\documentclass[seceq,preprint]{ptptex}
\usepackage{wrapft}
\usepackage{graphicx}

\newcommand{\ket}[1]{| {#1} \rangle}     
\newcommand{\bbra}[1]{\langle\!\langle {#1} |}     
\newcommand{\kket}[1]{| {#1} \rangle\!\rangle}     
\newcommand{\rket}[1]{| {#1} )}     
\newcommand{\dket}[1]{|\!| {#1} \rangle}     
\newcommand{\maru}[1]{\stackrel{\tiny\circ} {#1}} 
\newcommand{\wtilde}[1]{\widetilde{#1}} 

\newcommand{\kbar}{k \kern -0.5em\raise 0.6ex \hbox{--}}
\newcommand{\vect}[1]{{\overrightarrow{#1}}}

\def\<{\langle}
\def\>{\rangle}
\def\bsub{\begin{subequations}}
\def\esub{\end{subequations}}
\def\beqn{\begin{eqnarray}}
\def\eeqn{\end{eqnarray}}
\def\beq{\begin{equation}}
\def\eeq{\end{equation}}
\def\b{\begin{equation}}

\title{
The Lipkin Model in Many-Fermion System as\\
an Example of 
the 
$su(1,1)\otimes su(1,1)$-Algebraic Model
}

\author{
Constan\c{c}a {\sc Provid\^encia},$^{1}$
Jo\~ao da {\sc Provid\^encia},$^{1}$\\
Yasuhiko {\sc Tsue}$^{2}$ 
and Masatoshi {\sc Yamamura}$^{3}$
}

\inst{
$^{1}$Departamento de F\'\i sica, Universidade de Coimbra, P-3004-516 Coimbra, 
Portugal\\
$^2$Physics Division, Faculty of Science, Kochi University, Kochi 780-8520, 
Japan\\
$^{3}$Faculty of Engineering, Kansai University, Suita 564-8680, Japan
}

\recdate{
\today
}

\abst{
Following the idea, recently, proposed by the present authors for the 
two-level pairing model, the Lipkin model is reexamined. 
It is a natural generalization of the method already developed by the 
present authors with Kuriyama. 
This model is a schematic model for many-fermion system and 
obeys the $su(2)$-algebra. 
It is shown that the use of the Schwinger boson representation, the model 
is expressed in terms of the $su(1,1)\otimes su(1,1)$-algebra and 
with the aid of the MYT mapping method, it is disguised from the orginal 
form in terms of the Holstein-Primakoff representation. 
Further, under various coherent states, the classical counterparts are 
derived. 
It is concluded that the Lipkin model can be treated in the common ring 
as that of the two-level pairing model. 
}
\begin{document}
\maketitle

\section{Introduction}

The Lipkin model was originally proposed by Lipkin, Meshkov and Glick\cite{1} 
and has been widely used to test several kinds of many-body theories 
of strongly interacting fermion systems, for instance, the time-dependent 
Hartree-Fock method,\cite{2} finite temperature dynamics,\cite{3,4} 
collective dynamics of many-fermion system,\cite{5,6} 
phase transitions, spin tunnelling,\cite{7} etc. 
This statement can be found in the paper by the present authors with 
Kuriyama.\cite{8} 
Further, they have investigated the Lipkin model for thermal effect 
in Ref. \citen{9}. 
With the aim of participating in these researches under a certain new 
viewpoint for the Lipkin model, Ref. \citen{8} was published. 
On the basis of a new boson realization 
proposed by the present authors with Kuriyama,\cite{10} various 
numerical results were presented in Ref. \citen{8}

It is well known that, as was already mentioned, the Lipkin model is a 
schematic model for understanding the dynamics of many-fermion systems 
and it obeys the $su(2)$-algebra. 
The interaction term in the Hamiltonian is expressed in terms of the quadratic 
form with respect to the raising and the lowering operator in the 
$su(2)$-generators. 
On the other hand, we know that the $su(2)$-generators can be expressed 
in terms of the bilinear forms of two kinds of boson creation 
and annihilation operator, i.e., the Schwinger boson representation. 
Therefore, we can apply the Schwinger boson representation to the 
Lipkin model and the interaction term can be re-formed in terms of the 
product of the boson-pair creation and annihilation. 
The boson pairs and the boson number operator form the $su(1,1)$-algebra, 
and then, the Lipkin model is formulated in the frame of the 
$su(1,1)\otimes su(1,1)$-algebra. 
With the aid of the MYT mapping method,\cite{11} we can transcribe the 
system in the Holstein-Primakoff representation for the 
$su(1,1)$-algebra, and in Refs. \citen{8} and \citen{10}, 
a certain special case 
was treated. 

In response to the above situation, very recently, the present authors 
proposed a viewpoint that the two-level pairing model in many-fermion systems 
is also described in terms of the 
$su(1,1)\otimes su(1,1)$-algebra.\cite{12,13} 
Hereafter, Ref. \citen{13} is referred to as (I). 
The two-level pairing 
model is a kind of the $su(2)\otimes su(2)$-algebraic model, and 
with the aid of four kinds of bosons, we can treat this model 
in the Schwinger boson representation. 
In this sense, also, by applying an idea similar to that adopted in 
Refs. \citen{8} and \citen{10}, we can show that the present system is 
reduced to that obeying the $su(1,1)\otimes su(1,1)$-algebra.\cite{12}

A main aim of the present paper is to show again that the Lipkin model is a 
kind of the $su(1,1)\otimes su(1,1)$-algebraic model. 
Of course, the basic idea is similar to that adopted in Ref.\citen{12}. 
In contrast to the case of the two-level pairing model, the 
$su(1,1)$-generators in the Lipkin model consist of boson-pairs and 
boson number operators. 
Therefore, the structures of the $su(1,1)$-algebra are different from 
each other. 
The present case is characterized by the quantum numbers 
$t=1/4$ and $3/4$ and in Refs. \citen{8} and \citen{10}, 
only the case $t=1/4$ was treated. 
As was shown in Ref. \citen{12}, the two-level pairing model is characterized 
by the quantum numbers $t=1/2,\ 1,\ 3/2,\cdots$. 
This difference is not so important for applying the MYT boson mapping method 
to obtain the Holstein-Primakoff representation. 
However, it influences us into constructing boson coherent states which lead 
to the classical counterpart of the models. 
The two-level pairing and the Lipkin model are expressed in terms of four 
and two kinds of bosons, respectively. 
Therefore, compared with the two-level pairing model, a possibility to 
construct boson coherent state in the Lipkin model is limited and a 
special device is necessary. 
Under the above-mentioned scheme, we arrive at the following conclusion: 
We can describe the Lipkin model and the two-level pairing 
model in common ring. 
Of course, the comparison with the result based on the 
$su(2)\otimes su(2)$-coherent state is necessary. 

In \S 2, the Lipkin model in many-fermion systems is recapitulated with 
some new aspects and in \S 3, the Schwinger boson representation is 
described. 
Section 4 is devoted to re-forming the Lipkin model in the Schwinger 
boson representation. 
Of course, the MYT mapping method and a certain boson coherent state, which is 
called the $su(1,1)\otimes su(1,1)$-coherent state, 
play a central role and the model 
can be expressed in terms of two kinds of new bosons. 
In \S 5, the model is formulated in the frame of one kind of boson. 
Finally, in \S 6, the comparison with the result 
based on the $su(2)\otimes su(2)$-coherent state is sketched. 
In Appendix A, a possible idea for constructing the orthogonal 
set for the Lipkin model in the fermion space is demonstrated. 
In Appendix B, the boson-pair coherent state and a certain new boson 
coherent state are formulated.

\section{The original form of the Lipkin model in the fermion space}

The model discussed in this paper is the Lipkin model, which enables 
us to obtain a schematic understanding of the problems mentioned in the 
opening paragraph of \S 1. 
The framework of this model consists of two single-particle levels 
with the same degeneracies $\Omega=2j+1$ 
($\Omega$; even integer, $j$; half-integer), in which $N$ fermions 
are moving a certain type of the interaction. 
We specify the upper and the lower level by $\sigma=+$ and 
$\sigma=-$, respectively. 
The Hamiltonian ${\hat {\cal H}}$ is expressed in the form 
\begin{equation}\label{2-1}
{\hat {\cal H}}=\epsilon{\hat {\cal S}}_0-(G/4)
\left[\left({\hat {\cal S}}_+\right)^2+
\left({\hat {\cal S}}_-\right)^2\right] \ . 
\end{equation}
Here, ${\hat {\cal S}}_{\pm,0}$ are defined as 
\beqn\label{2-2}
& &{\hat {\cal S}}_+=\hbar\sum_{m=-j}^{j}{\hat c}_m^*(+){\hat c}_m(-) \ , 
\qquad
{\hat {\cal S}}_-=\hbar\sum_{m=-j}^{j}{\hat c}_m^*(-){\hat c}_m(+) \ , 
\nonumber\\
& &
{\hat {\cal S}}_0=(\hbar/2)\sum_{m=-j}^{j}({\hat c}_m^*(+){\hat c}_m(+) 
-{\hat c}_m^*(-){\hat c}_m(-))\ . 
\eeqn
The parameters $\epsilon$ and $G$ denote the difference of the single-particle 
energies between the two levels and the strength of the interaction, 
respectively. 
The set $({\hat c}_m(\sigma),{\hat c}_m^*(\sigma))$ represents 
fermion operator in the single-particle state $(\sigma, j, m; 
m=-j,-j+1,\cdots , j-1, j)$. 
Total fermion number operator ${\hat {\cal N}}$ is given as 
\begin{equation}\label{2-3}
{\hat {\cal N}}=\sum_{m=-j}^{j}
({\hat c}_m^*(+){\hat c}_m(+)+{\hat c}_m^*(-){\hat c}_m(-))\ . 
\end{equation}

We know that the set $({\hat {\cal S}}_{\pm,0})$ obeys the $su(2)$-algebra, 
and then, the Hamiltonian ${\hat {\cal H}}$ is diagonalized in the 
frame of the orthogonal set obtained in the well-known procedure. 
In this sense, it may be not necessary to repeat the discussion on the 
eigenvalue problem for the Lipkin model. 
However, as was mentioned in \S 1, we will describe this model in the 
framework of the Schwinger boson representation and the quantum 
numbers characterizing this model formally do not connect with 
those appearing in this model as a many-fermion system 
we are discussing in this section. 
One of the typical examples may be total fermion number, 
which is not contained in the Schwinger boson representation. 
Therefore, inevitably, focusing on the quantum numbers, 
we have to repeat the discussion on the eigenvalue problem.

If the state with the minimum weight, which we denote $\rket{m}$, 
is derived, we can construct the orthogonal set by operating the 
raising operator ${\hat {\cal S}}_+$ successively on $\rket{m}$. 
In the present case, $\rket{m}$ should satisfy the condition 
\bsub\label{2-4}
\beqn
& &{\hat {\cal S}}_-\rket{m}=0 \ , 
\label{2-4a}\\
& &{\hat {\cal S}}_0\rket{m}=-S\rket{m} \ , \quad
(S=\hbar s; s=0,1/2,1,\cdots )
\label{2-4b}\\
& &{\hat {\cal N}}\rket{m}=N\rket{m} \ , \quad\ \ 
(N=0,1, 2,\cdots , 2\Omega)
\label{2-4c}
\eeqn
\esub
In order to understand the structure of $\rket{m}$ obeying the 
condition (\ref{2-4}), first, we set up the following form of $\rket{m}$: 
\begin{equation}\label{2-5}
\rket{m}={\hat X}_\nu^*{\hat Y}_s^*\rket{0} \ .
\end{equation}
Here, ${\hat X}_\nu^*$ and ${\hat Y}_s^*$ are operators which should satisfy 
the condition 
\bsub\label{2-6}
\beqn
& &[\ {\hat {\cal S}}_{\pm}\ , \ {\hat X}_\nu^*\ ]=0 \ , \qquad
[\ {\hat {\cal S}}_{0}\ , \ {\hat X}_\nu^*\ ]=0 \ , \nonumber\\
& &[\ {\hat {\cal N}}\ , \ {\hat X}_\nu^*\ ]=2\nu{\hat X}_\nu^* \ , 
\quad (\nu; {\rm positive\ integer})\ , 
\label{2-6a}\\
& &[\ {\hat {\cal S}}_{-}\ , \ {\hat Y}_s^*\ ]=0 \ , \nonumber\\
& &[\ {\hat {\cal S}}_{0}\ , \ {\hat Y}_s^*\ ]=-\hbar s{\hat Y}_s^* \ , 
\nonumber\\
& &[\ {\hat {\cal N}}\ , \ {\hat Y}_s^*\ ]=2s{\hat Y}_s^* \ , 
\label{2-6b}\\
& &{\hat X}_\nu{\hat Y}_s^*\rket{0}=0 \ . 
\label{2-6c}
\eeqn
\esub
If ${\hat X}_\nu^*$ and ${\hat Y}_s^*$ obey the condition (\ref{2-6}), 
we can prove easily that the form (\ref{2-5}) satisfies the condition 
(\ref{2-4}). 
In order to investigate ${\hat X}_\nu^*$ and ${\hat Y}_s^*$ in rather 
concrete form, we note 
\beqn\label{2-7}
& &[\ {\hat {\cal S}}_{\pm}\ , \ {\hat c}_m^*(\pm) \ ]=0 \ , \qquad
[\ {\hat {\cal S}}_{\mp}\ , \ {\hat c}_m^*(\pm) \ ]=\hbar{\hat c}_m^*(\mp) 
\ , \nonumber\\
& &[\ {\hat {\cal S}}_0\ , \ {\hat c}_m^*(\pm) \ ]=
\mp (\hbar/2){\hat c}_m^*(\pm) \ , \qquad
[\ {\hat {\cal N}}\ , \ {\hat c}_m^*(\pm) \ ]={\hat c}_m^*(\pm) \ . 
\eeqn
We can see that the operator ${\hat c}_m^*(\pm)$ is spinor with respect 
to $({\hat {\cal S}}_{\pm,0})$. 
With the use of the relation (\ref{2-7}), we are permitted to set up 
\bsub\label{2-8}
\beqn
& &{\hat X}_\nu^*=\sum_{(k,l)}
c(k_1,\cdots ,k_\nu; l_1,\cdots ,l_\nu){\hat c}_{k_1}^*(+)\cdots
{\hat c}_{k_\nu}^*(+){\hat c}_{l_1}^*(-)\cdots {\hat c}_{l_\nu}^*(-) \ ,
\label{2-8a}\\
& &{\hat Y}_s^*={\hat c}_{m_1}^*(-)\cdots {\hat c}_{m_{2s}}^*(-) \ . 
\label{2-8b}
\eeqn
\esub
Here, $c(k_1,\cdots ,k_\nu; l_1,\cdots ,l_\nu)$ may be determined by 
appropriate method, which is sketched in Appendix A, but, in this paper, 
the explicit form is not necessary. 
Clearly, there exists the relation 
\begin{equation}\label{2-9}
\nu=N/2-s\ . 
\end{equation}
Then, we have the restriction
\beq\label{2-10}
0 \leq \nu \leq \Omega\ , \qquad 0 \leq \nu +2s \leq \Omega \ .
\end{equation}
The relation (\ref{2-10}) gives us $0\leq \nu \leq \Omega-2s$ 
and using the relation (\ref{2-9}), we have 
\beq\label{2-11}
0 \leq s \leq N/2 \ , \qquad 
0\leq s \leq \Omega-N/2\ .
\end{equation}
The relation (\ref{2-11}) is equivalent to 
\beqn\label{2-12}
& &{\rm if}\quad 0\leq N \leq \Omega \ , \qquad 0\leq s \leq N/2 \ , 
\nonumber\\
& &{\rm if}\quad \Omega\leq N \leq 2\Omega \ , 
\qquad 0\leq s \leq \Omega-N/2 \ . 
\eeqn
Noting that $\nu$ is a positive integer, the relation (\ref{2-12}) 
can be expressed as follows: 
\beqn\label{2-13}
& &{\rm if}\quad N={\rm even\ and}\ 
0\leq N \leq \Omega \ , \qquad s=N/2,\ N/2-1,\ \cdots ,\ 1,\ 0 \ ,
\nonumber\\
& &{\rm if}\quad N={\rm even\ and}\ 
\Omega\leq N \leq 2\Omega \ , \qquad s=\Omega-N/2,\ \Omega-N/2-1,\ 
\cdots ,\ 1,\ 0 \ ,
\nonumber\\
& &{\rm if}\quad N={\rm odd\ and}\ 
0 < N < \Omega \ , \qquad s=N/2,\ N/2-1,\ \cdots ,\ 3/2,\ 1/2 \ ,
\nonumber\\
& &{\rm if}\quad N={\rm odd\ and}\ 
\Omega < N < 2\Omega \ , \qquad s=\Omega-N/2,\ \Omega-N/2-1,\ 
\cdots ,\ 3/2,\ 1/2 \ .\nonumber\\
& &
\eeqn
The relation (\ref{2-9}) tells us that $2\nu$ denotes the seniority 
number, and usually, the Lipkin model is treated in the case 
$N=\Omega$ with $s=\Omega/2$ ($\nu=0$). 
The above means that the lower level is fully occupied by $\Omega$ fermions 
and the upper is empty if the interaction is switched off.

As a final discussion in this section, we will contact with the problem how 
to construct the orthogonal set. 
Since $\rket{m}$ satisfies the conditions (\ref{2-4a}) and (\ref{2-4b}), 
formally, 
$({\hat {\cal S}}_+)^{\sigma}\rket{m}\ (\sigma=s+s_0)$ gives us the state with 
the eigenvalue $(s,s_0; s_0=-s,-s+1,\cdots s-1,s)$. 
However, further consideration is necessary. 
The Hamiltonian (\ref{2-1}) consists of ${\hat {\cal S}}_0$, 
${\hat {\cal S}}_+^2$ and ${\hat {\cal S}}_-^2$. 
Therefore, the matrix elements of ${\hat {\cal H}}$ between two states with 
$\Delta s_0=\pm 1$ always vanish. 
This means that the whole space specified by a given value of $s$ 
is divided into two groups, i.e., 
$\{({\hat {\cal S}}_+)^{2n}\rket{m}\}$ 
and $\{({\hat {\cal S}}_+)^{2n+1}\rket{m}\}$. 
Here, $n$ denotes $n=0,1,2,\cdots $ and the normalization constants 
are omitted. 
For the group $\{({\hat {\cal S}}_+)^{2n}\rket{m}\}$, $n$ is restricted to 
\bsub\label{2-13sub}
\beq\label{2-13a}
-s \leq -s+2n \leq s\ , \qquad {\rm i.e.,}\qquad 0\leq n \leq s \ . 
\end{equation}
On the other hand, for the group $\{({\hat {\cal S}}_+)^{2n+1}\rket{m}\}$, 
we have 
\beq\label{2-13b}
-s+1 \leq -s+2n+1 \leq s\ , \qquad {\rm i.e.,}\qquad 0\leq n \leq s-1/2 \ . 
\end{equation}
\esub
Noting that $s$ is integer or half-integer, the orthogonal set for 
diagonalizing ${\hat {\cal H}}$ is classified into four groups: 
\bsub\label{2-14}
\beqn
& &{\rm (i)}\quad 
\{({\hat {\cal S}}_+)^{2n}\rket{m} ; s={\rm integer},\ n=0,1,2,\cdots ,s\} \ , 
\label{2-14a}\\
& &{\rm (ii)}\quad 
\{({\hat {\cal S}}_+)^{2n+1}\rket{m} ; s={\rm integer},\ n=0,1,2,\cdots ,s-1\} 
\ , 
\label{2-14b}\\
& &{\rm (iii)}\quad 
\{({\hat {\cal S}}_+)^{2n}\rket{m} ; s=\hbox{\rm half-integer},\ 
n=0,1,2,\cdots ,s-1/2\} \ , 
\label{2-14c}\\
& &{\rm (iv)}\quad 
\{({\hat {\cal S}}_+)^{2n+1}\rket{m} ; s=\hbox{\rm half-integer},\ 
n=0,1,2,\cdots ,s-1/2\} \ . 
\label{2-14d}
\eeqn
\esub
Through the above-mentioned procedure, we can prepare the orthogonal set for 
the original form of the Lipkin model.

\section{The Lipkin model in the Schwinger boson representation}

It is well known that the $su(2)$-generators in the Schwinger boson 
representation are expressed in terms of two kinds of bosons 
$({\hat b}_+,{\hat b}_+^*)$ and $({\hat b}_-,{\hat b}_-^*)$: 
\beq\label{3-1}
{\wtilde S}_+=\hbar{\hat b}_+^*{\hat b}_-\ , \qquad
{\wtilde S}_-=\hbar{\hat b}_-^*{\hat b}_+\ , \qquad
{\wtilde S}_0=(\hbar/2)({\hat b}_+^*{\hat b}_+
-{\hat b}_-^*{\hat b}_-)\ .
\end{equation}
The operator expressing the magnitude of the $su(2)$-spin is given 
in the form 
\beq\label{3-2}
{\wtilde S}=(\hbar/2)({\hat b}_+^*{\hat b}_+
+{\hat b}_-^*{\hat b}_-)\ .
\end{equation}
In the fermion space, it may be impossible to describe such a simple 
expression. 
The state with the minimum weight, which we denote as $\kket{m}$, 
is given as 
\beq\label{3-3}
\kket{m}=\left(\sqrt{(2s)!}\right)^{-1}({\hat b}_-^*)^{2s}\kket{0}\ . 
\qquad (s=0, 1/2, 1,\cdots )
\end{equation}
The state $\kket{m}$ satisfies 
\bsub\label{3-4}
\beqn
& &{\wtilde S}_-\kket{m}=0 \ , 
\label{3-4a}\\
& &{\wtilde S}_0\kket{m}=-S\kket{m}\ , \qquad (S=\hbar s) \ .
\label{3-4b}
\eeqn
\esub
In the Schwinger boson representation, there does not exist any operator 
which corresponds to ${\hat {\cal N}}$, and then, we cannot set up the 
relation such as (\ref{2-4c}). 
The successive operation of ${\wtilde S}_+$ on the state $\kket{m}$ gives us 
the orthogonal set in the Schwinger boson representation. 
Later, we will discuss this problem.

The Hamiltonian ${\wtilde H}$ which corresponds to ${\hat {\cal H}}$ 
shown in the form (\ref{2-1}) is, of course, expressed as follows: 
\beq\label{3-5}
{\wtilde H}=\epsilon{\wtilde S}_0 -(G/4)
\left[\left({\wtilde S}_+\right)^2+\left({\wtilde S}_-\right)^2\right] \ .
\end{equation}
Any matrix element of ${\wtilde H}$ for the orthogonal set derived 
from $\kket{m}$ is of the same form as that of ${\hat {\cal H}}$ for the 
orthogonal set derived from $\rket{m}$. 
However, the original form of the Lipkin model is characterized by the 
quantities $\Omega$ and $N$, which give us the restriction 
(\ref{2-13}). 
On the other hand, the Lipkin model in the Schwinger boson representation 
does not contain such quantities. 
Then, introducing $\Omega$ and $N$ from the outside, 
we require the relation (\ref{2-13}) as the condition that the 
magnitude of the $su(2)$-spin $s$, the eigenvalue of ${\wtilde S}$ 
shown in the relation (\ref{3-2}), should satisfy in the Schwinger 
boson representation. 
Through this requirement, both frameworks connect with each other.

Now, let us construct the orthogonal set for diagonalizing ${\wtilde H}$ 
shown in the relation (\ref{3-5}). 
The Hamiltonian ${\wtilde H}$ also consists of ${\wtilde S}_0$, 
${\wtilde S}_+^2$ and ${\wtilde S}_-^2$. 
Therefore, the same argument as that given in \S 2 is possible. 
We can divide the whole space into two groups, 
$\{({\wtilde S}_+)^{2n}\kket{m}\}$ and $\{({\wtilde S}_+)^{2n+1}\kket{m}\}$. 
Except the normalization constants, both states can be rewritten as 
\bsub\label{3-6}
\beqn
& &\left({\wtilde S}_+\right)^{2n}\kket{m}=
({\hat b}_+^*)^{2n}({\hat b}_-^*)^{2(s-n)}\kket{0} \ , 
\label{3-6a}\\
& &\left({\wtilde S}_+\right)^{2n+1}\kket{m}=
({\hat b}_+^*)^{2n+1}({\hat b}_-^*)^{2(s-n)-1}\kket{0} \ . 
\label{3-6b}
\eeqn
\esub
The form (\ref{3-6a}) gives us 
\bsub\label{3-7}
\beq\label{3-7a}
2n \geq 0 \ , \qquad 2(s-n) \geq 0 \ , \qquad{\rm i.e.,}\qquad
0\leq n\leq s \ .
\end{equation}
Also, the form (\ref{3-6b}) gives us 
\beq\label{3-7b}
2n+1 \geq 0 \ , \qquad 2(s-n)-1 \geq 0 \ , \qquad{\rm i.e.,}\qquad
0\leq n\leq s-1/2 \ .
\end{equation}
\esub
The relation (\ref{3-7}) is identical to the relation (\ref{2-13sub}). 
Under the same idea as that used in \S 2, the orthogonal set is also 
classified into four groups: 
\bsub\label{3-8}
\beqn
& &{\rm (i)}\quad 
\{({\wtilde S}_+)^{2n}\kket{m}=({\hat b}_+^{*2})^n({\hat b}_-^{*2})^{s-n}
\kket{0} ; \nonumber\\
& &\qquad\qquad
s={\rm integer},\ n=0,1,2,\cdots ,s\} \ , 
\label{3-8a}\\
& &{\rm (ii)}\quad 
\{({\wtilde S}_+)^{2n+1}\kket{m}=({\hat b}_+^{*2})^n({\hat b}_-^{*2})^{s-n-1}
\cdot{\hat b}_+^*{\hat b}_-^*\kket{0} ; \nonumber\\
& &\qquad\qquad
s={\rm integer},\ n=0,1,2,\cdots ,s-1\} \ , 
\label{3-8b}\\
& &{\rm (iii)}\quad 
\{({\wtilde S}_+)^{2n}\kket{m}=({\hat b}_+^{*2})^n({\hat b}_-^{*2})^{s-n-1/2}
\cdot{\hat b}_-^*\kket{0} ; \nonumber\\
& &\qquad\qquad
s=\hbox{\rm half-integer},\ n=0,1,2,\cdots ,s-1/2\} \ , 
\label{3-8c}\\
& &{\rm (iv)}\quad 
\{({\wtilde S}_+)^{2n+1}\kket{m}=({\hat b}_+^{*2})^n({\hat b}_-^{*2})^{s-n-1/2}
\cdot{\hat b}_+^*\kket{0} ; \nonumber\\
& &\qquad\qquad
s=\hbox{\rm half-integer},\ n=0,1,2,\cdots ,s-1/2\} \ . 
\label{3-8d}
\eeqn
\esub

From the relation (\ref{3-8}), we can learn that for a given value of $s$, 
the maximum number of the operation of ${\wtilde S}_+^2$ is fixed. 
We also mentioned that the quantities $\Omega$ and $N$ are introduced from 
the outside and they should obey the condition (\ref{2-13}). 
However, it may be convenient for the practical aim to rewrite the 
condition (\ref{2-13}) in the form so as to be able to know the 
values of $N$ permitted for a given value of $s$. 
This can be done in the form 
\bsub\label{3-9}
\beqn
& &{\rm (a)}\ 
{\rm if}\ s={\rm integer},\quad N=2s,2s+2,\cdots , \Omega, \nonumber\\
& &\qquad\qquad\qquad\qquad\qquad\qquad\qquad
2\Omega-2s, 
2\Omega-2s-2,\cdots ,2\Omega\ , \quad
\label{3-9a}\\
& &{\rm (b)}\ 
{\rm if}\ s=\hbox{\rm half-integer},
\quad N=2s,2s+2,\cdots ,\Omega-1, \nonumber\\
& &\qquad\qquad\qquad\qquad\qquad\qquad\qquad
2\Omega-2s, 
2\Omega-2s-2,\cdots ,2\Omega-1\ . 
\label{3-9b}
\eeqn
\esub
The above is the outline of the Lipkin model in the Schwinger boson 
representation.

\section{Reformulation in the form of the $su(1,1)\otimes su(1,1)$-algebra}

In Refs. \citen{12} and \citen{13}, we showed that the two-level pairing 
model 
was re-formed in terms of the $su(1,1)\otimes su(1,1)$-algebraic model 
in the Schwinger boson representation. 
The present authors, with Kuriyama, already showed that the Lipkin model 
can be also re-formed in terms of the $su(1,1)\otimes su(1,1)$-algebraic 
model in the Schwinger boson representation.\cite{8,10} 
In this section, we recapitulate this re-formation in the notations used 
in Refs. \citen{12} and \citen{13}. 
Of course, the formalism given in Refs. \citen{8} and \citen{10} is 
supplemented with newly added features. 

First, we note the following re-form: 
\bsub\label{4-1}
\beqn
& &\left({\wtilde S}_+\right)^2=4(\hbar/2){\hat b}_+^{*2}\cdot
(\hbar/2){\hat b}_-^2 \ , 
\label{4-1a}\\
& &\left({\wtilde S}_-\right)^2=4(\hbar/2){\hat b}_-^{*2}\cdot
(\hbar/2){\hat b}_+^2 \ , 
\label{4-1b}\\
& &{\wtilde S}_0=(\hbar/4+(\hbar/2){\hat b}_+^*{\hat b}_+)-
(\hbar/4+(\hbar/2){\hat b}_-^*{\hat b}_-) \ . 
\label{4-1c}
\eeqn
\esub
Then, we define the  boson-pairs in the set $({\wtilde T}_{\pm,0}(\sigma))$: 
\beqn\label{4-2}
& &{\wtilde T}_+(\sigma)=(\hbar/2){\hat b}_{\sigma}^{*2} \ , \quad
{\wtilde T}_-(\sigma)=(\hbar/2){\hat b}_{\sigma}^{2} \ , \quad
{\wtilde T}_0(\sigma)=\hbar/4+(\hbar/2){\hat b}_{\sigma}^{*}{\hat b}_{\sigma} 
\ . \quad (\sigma=\pm)\nonumber\\
& &
\eeqn
It is easily verified that the set $({\wtilde T}_{\pm,0}(\sigma))$ 
obeys the $su(1,1)$-algebra. 
With the use of the set $({\wtilde T}_{\pm,0}(\sigma);\sigma=\pm)$, 
the Hamiltonian (\ref{3-5}) can be rewritten as 
\beq\label{4-3}
{\wtilde H}=\epsilon({\wtilde T}_0(+)-{\wtilde T}_0(-))
-G\left({\wtilde T}_+(+){\wtilde T}_-(-)
+{\wtilde T}_+(-){\wtilde T}_-(+)\right) \ . 
\end{equation}
The form (\ref{4-3}) tells that the Lipkin model is a possible model 
obeying the $su(1,1)\otimes su(1,1)$-algebra. 

Our concern is to compare the Lipkin model with the two-level pairing model 
in the framework of the $su(1,1)\otimes su(1,1)$-algebra. 
For this purpose, further, we re-form the present frame. 
The state with the minimum weight, $\kket{m(\sigma)}$, for the algebra 
$({\wtilde T}_{\pm,0}(\sigma))$ is determined under the condition 
\bsub\label{4-4}
\beqn
& &{\wtilde T}_-(\sigma)\kket{m(\sigma)}=0 \ , 
\label{4-4a}\\
& &{\wtilde T}_0(\sigma)\kket{m(\sigma)}=T(\sigma)\kket{m(\sigma)} \ , 
\qquad (T(\sigma)=\hbar t_{\sigma}) \ .
\label{4-4b}
\eeqn
\esub
The condition (\ref{4-4}), with the definition (\ref{4-2}), gives 
\begin{equation}\label{4-5}
\kket{m(\sigma)}=({\hat b}_{\sigma}^*)^{2t_{\sigma}-1/2}\kket{0}=
\begin{cases} \kket{0} \ , \qquad {\rm for}\quad t_\sigma=1/4 \ , \\
{\hat b}_{\sigma}^*\kket{0} \ , \quad{\rm for}\quad t_{\sigma}=3/4 \ . 
\end{cases}
\end{equation}
It should be noted that there does not exist any other type. 
Then, the state with the minimum weight in the present 
$su(1,1)\otimes su(1,1)$-algebra is specified by $(t_+,t_-)$ and 
given in the form 
\beq\label{4-6}
\kket{t_+,t_-}=({\hat b}_+^*)^{2t_+-1/2}({\hat b}_-^*)^{2t_--1/2}\kket{0} \ .
\end{equation}
Then, we can construct the orthogonal set by operating 
${\wtilde T}_+(+)$ and ${\wtilde T}_+(-)$ successively in the form 
\beqn\label{4-9}
& &\kket{\kappa_+,\kappa_-;t_+,t_-}=
\left(\sqrt{(2\kappa_++2t_+-1/2)!(2\kappa_-+2t_--1/2)!}\right)^{-1}\nonumber\\
& &\qquad\qquad\qquad\qquad\qquad
\times
({\hat b}_+^*)^{2\kappa_+}({\hat b}_-^*)^{2\kappa_-}\kket{t_+,t_-} \ , 
\nonumber\\
& &\qquad 
\kappa_+,\ \kappa_-=0,1,2,\cdots .
\eeqn
It may be interesting to compare the state (\ref{4-9}) with the form 
(\ref{3-8}). 
We can see that both are identical with each other under the 
following correspondence: 
\bsub\label{4-10}
\beqn
& &{\rm (i)}\ \ t_+=1/4 \ , \quad t_-=1/4\ , \quad \kappa_+=n \ , \quad 
\kappa_-=s-n \ , 
\label{4-10a}\\
& &{\rm (ii)}\ \ t_+=3/4 \ , \quad t_-=3/4\ , \quad \kappa_+=n \ , \quad 
\kappa_-=s-n-1 \ , 
\label{4-10b}\\
& &{\rm (iii)}\ \ t_+=1/4 \ , \quad t_-=3/4\ , \quad \kappa_+=n \ , \quad 
\kappa_-=s-n-1/2 \ , 
\label{4-10c}\\
& &{\rm (iv)}\ \ t_+=3/4 \ , \quad t_-=1/4\ , \quad \kappa_+=n \ , \quad 
\kappa_-=s-n-1/2 \ . 
\label{4-10d}
\eeqn
\esub
We know that the quantum number $(t_+,t_-)$ plays a role of classifying the 
four cases (i)$\sim$(iv).

Next, as was done in Ref. \citen{13}, we apply the MYT mapping method\cite{11} 
to the present system. 
For this purpose, we prepare the other boson space constructed by new 
boson $({\hat c}_{\sigma}, {\hat c}_{\sigma}^*; \sigma=\pm)$. 
The orthogonal set is given as 
\beq\label{4-11}
\dket{\kappa_+,\kappa_-}=\left(\sqrt{\kappa_+!\kappa_-!}\right)^{-1}
({\hat c}_+^*)^{\kappa_+}({\hat c}_-^*)^{\kappa_-}\dket{0} \ . 
\end{equation}
Then, the MYT mapping operator $\maru{U}$ is defined in the form 
\beq\label{4-12}
\maru{U}=\sum_{\kappa_+,\kappa_-}\dket{\kappa_+,\kappa_-}
\bbra{\kappa_+,\kappa_-;t_+,t_-} \ . 
\end{equation}
With the use of $\maru{U}$, we have the following relations: 
\beqn
& &\maru{U}\left({\wtilde T}_0(+)-{\wtilde T}_0(-)\right)\maru{U}^{\dagger}
=T(+)-T(-)+\hbar{\hat c}_+^*{\hat c}_+-\hbar{\hat c}_-^*{\hat c}_- \ , 
\label{4-13}\\
& &\maru{U}{\wtilde T}_+(+){\wtilde T}_-(-)\maru{U}^{\dagger}
=\sqrt{2T(-)+\hbar{\hat c}_-^*{\hat c}_-}\cdot\hbar{\hat c}_+^*{\hat c}_-
\cdot\sqrt{2T(+)+\hbar{\hat c}_+^*{\hat c}_+} \ , 
\label{4-14}\\
& &T(+)=\hbar t_+ \ , \qquad T(-)=\hbar t_- \ . 
\label{4-15}
\eeqn
With the help of the forms (\ref{4-13}) and (\ref{4-14}), the Hamiltonian 
(\ref{4-3}) can be mapped to the following: 
\beqn\label{4-16}
\maru{H}&=&
\epsilon(T(+)-T(-))+\epsilon(\hbar{\hat c}_+^*{\hat c}_+
-\hbar{\hat c}_-^*{\hat c}_-)\nonumber\\
& &-G\biggl(
\sqrt{2T(-)+\hbar{\hat c}_-^*{\hat c}_-}\cdot \hbar{\hat c}_+^*{\hat c}_-\cdot
\sqrt{2T(+)+\hbar{\hat c}_+^*{\hat c}_+} \nonumber\\
& &\qquad\quad
+\sqrt{2T(+)+\hbar{\hat c}_+^*{\hat c}_+}\cdot\hbar{\hat c}_-^*{\hat c}_+\cdot
\sqrt{2T(-)+\hbar{\hat c}_-^*{\hat c}_-}\biggl) \ .
\eeqn
The magnitude of the $su(2)$-spin, ${\wtilde S}$, shown in the relation 
(\ref{3-2}) is mapped to 
\beq\label{4-17}
\maru{S}=\maru{U}{\wtilde S}\maru{U}^{\dagger}
=T(+)+T(-)-\hbar/2+2\cdot(\hbar/2)({\hat c}_+^*{\hat c}_+
+{\hat c}_-^*{\hat c}_-) \ . 
\end{equation}
Thus, we can construct the Lipkin model in the framework of the 
$su(1,1)\otimes su(1,1)$-algebra in the Holstein-Primakoff representation. 

Comparison of the form (\ref{4-16}) with the Hamiltonian (I$\cdot$13) is 
interesting. 
Except a certain term related to $\hbar{\hat c}_+^*{\hat c}_+$ and 
$\hbar{\hat c}_-^*{\hat c}_-$, both coincide with each other. 
Of course, the form (I$\cdot$13) is characterized by 
\beq\label{4-18}
T(\sigma)=\hbar/2, \hbar, 3\hbar/2,\cdots \ .
\end{equation}
The present case is characterized by $T(\sigma)=\hbar/4$ and $3\hbar/4$. 
From the above argument, we can conclude that the two-level pairing 
and the Lipkin model can be treated in the common ring. 
The Hamiltonian is expressed as 
\beqn\label{4-19}
\maru{H}&=&
\epsilon(T(+)-T(-))+\epsilon(\hbar{\hat c}_+^*{\hat c}_+
-\hbar{\hat c}_-^*{\hat c}_-)\nonumber\\
& &-2g(T(+)\cdot\hbar{\hat c}_+^*{\hat c}_+ 
+ T(-)\cdot\hbar{\hat c}_-^*{\hat c}_-)
-2g\cdot\hbar{\hat c}_+^*{\hat c}_+\cdot \hbar{\hat c}_-^*{\hat c}_- 
\nonumber\\
& &-G\biggl(
\sqrt{2T(-)+\hbar{\hat c}_-^*{\hat c}_-}\cdot \hbar{\hat c}_+^*{\hat c}_-\cdot
\sqrt{2T(+)+\hbar{\hat c}_+^*{\hat c}_+} \nonumber\\
& &\qquad\quad
+\sqrt{2T(+)+\hbar{\hat c}_+^*{\hat c}_+}\cdot\hbar{\hat c}_-^*{\hat c}_+\cdot
\sqrt{2T(-)+\hbar{\hat c}_-^*{\hat c}_-}\biggl) \ .
\eeqn
The cases $g=G$ and $g=0$ correspond to the two-level pairing and the 
Lipkin model, respectively. 
As was shown in (I), the Hamiltonian (\ref{4-19}) can be treated in the 
$su(2)$-algebra. 

Finally, we will show that a certain wave packet defined in the form 
(\ref{4-23}), which we call the $su(1,1)\otimes su(1,1)$-coherent state, 
presents 
a classical counterpart of the Lipkin model re-formed in this section. 
First, extending the state (\ref{b10}), we introduce the following 
state: 
\beqn
& &\kket{c_0}=\left(\sqrt{U_+}\right)^{-1}
\exp\left(\frac{V_+}{\hbar U_+}{\wtilde T}_+(+)\right)\cdot
\left(\sqrt{U_-}\right)^{-1}
\exp\left(\frac{V_-}{\hbar U_-}{\wtilde T}_+(-)\right)\kket{0} \ , 
\label{4-20}\\
& &U_{\sigma}=\sqrt{1+|V_{\sigma}|^2} \ . \qquad (\sigma=\pm)
\label{4-21}
\eeqn
Here, $V_{\sigma}$ denotes a complex parameter. 
The state $\kket{c_0}$ is a vacuum of boson operator $({\hat \beta}_{\sigma},
{\hat \beta}_{\sigma}^*)$, i.e., ${\hat \beta}_{\sigma}\kket{c_0}=0$: 
\bsub\label{4-22}
\beqn
& &{\hat \beta}_{\sigma}=U_{\sigma}{\hat b}_{\sigma}
-V_{\sigma}{\hat b}_{\sigma}^* 
\ , \nonumber\\
& &{\hat \beta}_{\sigma}^*=-V_{\sigma}^*{\hat b}_{\sigma}
+U_{\sigma}{\hat b}_{\sigma}^* \ , 
\label{4-20a}\\
{\rm i.e.,}\qquad
& &{\hat b}_{\sigma}=U_{\sigma}{\hat \beta}_{\sigma}
+V_{\sigma}{\hat \beta}_{\sigma}^*\ , \nonumber\\
& &{\hat b}_{\sigma}^*=V_{\sigma}^*{\hat \beta}_{\sigma}
+U_{\sigma}{\hat \beta}_{\sigma}^* \ .
\label{4-20b}
\eeqn
\esub
With the use of $\kket{c_0}$ and ${\hat \beta}_{\sigma}^*$, we define the 
$su(1,1)\otimes su(1,1)$-coherent state $\kket{c^0}$ in the form 
\beq\label{4-23}
\kket{c^0}=\left(\sqrt{3/2-2t_+}+\sqrt{2t_+-1/2}{\hat \beta}_+^*\right)
\left(\sqrt{3/2-2t_-}+\sqrt{2t_--1/2}{\hat \beta}_-^*\right)\kket{c_0} \ . 
\end{equation}
The state (\ref{4-23}) is a possible extension of the state (\ref{b13}) 
to the case of the $su(1,1)\otimes su(1,1)$-algebra. 
By calculating the expectation value of ${\wtilde H}$ given in the relation 
(\ref{4-3}) for $\kket{c^0}$ and using various relations shown in Appendix, 
we have the following form: 
\beqn\label{4-24}
{H}^0&=&
\epsilon(T(+)-T(-))+\epsilon(\hbar{c}_+^*{c}_+
-\hbar{c}_-^*{c}_-)\nonumber\\
& &-G\biggl(
\sqrt{2T(-)+\hbar{c}_-^*{c}_-}\cdot \hbar{c}_+^*{c}_-\cdot
\sqrt{2T(+)+\hbar{c}_+^*{c}_+} \nonumber\\
& &\qquad\quad
+\sqrt{2T(+)+\hbar{c}_+^*{c}_+}\cdot\hbar{c}_-^*{c}_+\cdot
\sqrt{2T(-)+\hbar{c}_-^*{c}_-}\biggl) \ .
\eeqn
We can see that $H^0$ is equivalent to $\maru{H}$ shown in the 
relation (\ref{4-16}). 
Combining the above result with that derived in (I), 
we obtain the classical Hamiltonian of $\maru{H}$ given in the relation 
(\ref{4-19}). 
Of course, for the present case, $T(\sigma)$ should be used as 
$T(\sigma)=\hbar/4$ or $3\hbar/4$.

\section{Re-formation in terms of one kind of boson operator}

As was discussed in (I), the Hamiltonian (\ref{4-16}) is expressed 
in terms of the Schwinger boson representation for the $su(2)$-algebra: 
\beqn
& &{\maru M}_+=\hbar{\hat c}_+^*{\hat c}_- \ , \qquad
{\maru M}_-=\hbar{\hat c}_-^*{\hat c}_+ \ , \qquad
{\maru M}_0=(\hbar/2)({\hat c}_+^*{\hat c}_+-{\hat c}_-^*{\hat c}_-) \ , 
\label{5-1}\\
& &{\maru M}=(\hbar/2)({\hat c}_+^*{\hat c}_++{\hat c}_-^*{\hat c}_-) \ . 
\label{5-2}
\eeqn
Then, in the same idea as that shown in (I), we can re-form our present 
system in terms of one kind of boson. 
For example, we obtain the following form, which is the same as that shown 
in the relation (I$\cdot$30): 
\beqn\label{5-3}
& &\sqrt{2T(-)+\hbar{\hat c}_-^*{\hat c}_-}\cdot\hbar{\hat c}_+^*{\hat c}_-
\cdot\sqrt{2T(+)+\hbar{\hat c}_+^*{\hat c}_+} \nonumber\\
&\Longrightarrow& \sqrt{\hbar}{\hat c}^*
\sqrt{2T(+)+\hbar{\hat c}^*{\hat c}}\sqrt{2M-\hbar{\hat c}^*{\hat c}}
\sqrt{2T(-)+2M-\hbar-\hbar{\hat c}^*{\hat c}}\ . 
\eeqn
Using the canonical transformation 
$(\sqrt{\hbar}c_+=\sqrt{\hbar}ce^{-i\chi/2}, 
\sqrt{\hbar}c_-=
\sqrt{2M-\hbar c^*c}e^{-i\chi/2}$) adopted in (I), 
the classical counterpart obtained by the state $\kket{c^0}$ shown in 
the relation (\ref{4-23}) can be re-formed in terms of 
one kind of boson-type canonical variable. 
For example, we have 
\beqn\label{5-4}
& &\sqrt{2T(-)+\hbar{c}_-^*{c}_-}\cdot\hbar{c}_+^*{c}_-
\cdot\sqrt{2T(+)+\hbar{c}_+^*{c}_+} \nonumber\\
&\Longrightarrow& \sqrt{\hbar}{c}^*
\sqrt{2T(+)+\hbar{c}^*{c}}\sqrt{2M-\hbar{c}^*{c}}
\sqrt{2T(-)+2M-\hbar{c}^*{c}}\ . 
\eeqn
We can see that the form (\ref{5-4}) does not agree with that obtained by 
the $c$-number replacement of the relation (\ref{5-3}). 
This is also in the same situation as that in the case of the two-level 
pairing model. 
The form (\ref{5-4}) is obtained by the $su(1,1)\otimes su(1,1)$-coherent 
state and the other 
by the $su(2)\otimes su(1,1)$-coherent state. 
Then, we have a question: 
What type of the coherent state reproduces the $c$-number replaced 
form of the relation (\ref{5-3}) ? 

In order to give an answer to the above question, let us remember the case 
of the two-level pairing model. 
In this case, four kinds of bosons are divided into two groups. 
The $su(2)\otimes su(1,1)$-coherent 
state is constructed under the following idea: 
This state consists of the product of two parts. 
First consists of the exponential form for the raising operator of the 
$su(1,1)$-algebra and the second is expressed in terms of the 
Glauber form 
for two kinds of bosons. 
In the present case, the first part can be expressed in the same form as the 
above. 
However, if adopting the $su(2)\otimes su(2)$-coherent state for the second, 
we cannot derive the expected form. 
Under the above mentioned background, we adopt the following coherent state: 
\beq\label{5-5}
\kket{c_0'}=\left(\sqrt{U_+}\right)^{-1}\!\!
\exp\!\left(\frac{V_+}{\hbar U_+}{\wtilde T}_+(+)\right)
\left(\sqrt{U_-}\right)^{-1}\!\!\exp\!\left(\frac{\gamma_-}{|\gamma_-|}
\frac{V_-}{\hbar U_-}{\wtilde T}_+(-)+\frac{\sqrt{\gamma}_-}{U_-}{\hat b}_-^*
\right)\!\kket{0} \ . 
\end{equation}
Here, $V_+$ and $\gamma_-$ are complex, but $V_-$ is real and $U_+$ 
and $U_-$ are given as 
\beq\label{5-6}
U_+=\sqrt{1+|V_+|^2} \ , \qquad U_-=\sqrt{1+V_-^2}\ . 
\end{equation}
The detail properties of the state $\kket{c_0'}$ are discussed separately in 
Appendix B. 
Further, we define the state $\kket{{c^0}'}$ in the form 
\beq\label{5-7}
\kket{{c^0}'}=
\left(\sqrt{3/2-2t_+}+\sqrt{2t_+-1/2}{\hat \beta}_+^*\right)
\left(\sqrt{3/2-2t_-}+\sqrt{2t_--1/2}{\hat \beta}_-^*\right)\kket{c_0'} \ .
\end{equation}
Using the various relations shown in \S 4 and 
Appendix B and noting the relation $M=(\hbar/2)(c_+^*c_++c_-^*c_-)$, 
we get the following result for the Hamiltonian ${H^0}'$: 
\beqn\label{5-8}
{H^0}'&=&\bbra{{c^0}'}{\wtilde H}\kket{{c^0}'} \nonumber\\
&=&\epsilon(T(+)-T(-)-M)+2\epsilon\cdot \hbar c^* c \nonumber\\
& &-G\biggl( \sqrt{\hbar}c^*\cdot \sqrt{2T(+)+\hbar c^* c}
\sqrt{2M-\hbar c^* c}\sqrt{2T(-)+2M-\hbar-\hbar c^* c} \nonumber\\
& &\qquad +
\sqrt{2T(-)+2M-\hbar-\hbar c^* c}\sqrt{2M-\hbar c^* c}
\sqrt{2T(+)+\hbar c^* c}
\cdot \sqrt{\hbar}c \biggl) \ .\quad
\eeqn
Of course, the relation (\ref{5-3}) presents us the quantized form of 
${H^0}'$: 
\beqn\label{5-9}
{\maru{H'}}
&=&\epsilon(T(+)-T(-)-M)+2\epsilon\cdot \hbar {\hat c}^* {\hat c} \nonumber\\
& &-G\biggl( \sqrt{\hbar}{\hat c}^*\cdot \sqrt{2T(+)+\hbar{\hat c}^*{\hat c}}
\sqrt{2M-\hbar{\hat c}^*{\hat c}}\sqrt{2T(-)+2M-\hbar-\hbar{\hat c}^*{\hat c}} 
\nonumber\\
& &\qquad +
\sqrt{2T(-)+2M-\hbar-\hbar{\hat c}^*{\hat c}}\sqrt{2M-\hbar{\hat c}^*{\hat c}}
\sqrt{2T(+)+\hbar{\hat c}^*{\hat c}}
\cdot \sqrt{\hbar}{\hat c} \biggl) \ .\quad
\eeqn
The Hamiltonians ${H^0}'$ and $\maru{H'}$ contain the quantity $M$, 
which is an eigenvalue of $\maru{M}$ defined in the relation (\ref{5-2}). 
Further, we note the relation (\ref{4-17}) and $\maru{S}$ is related 
to ${\wtilde S}$ defined in the relation (\ref{3-2}). 
Therefore, we have the following relation: 
\beq\label{5-10}
S=T(+)+T(-)-\hbar/2+2M \ . 
\end{equation}
Through the relation (\ref{3-9}), $M$ connects with $\Omega$ and $N$ 
in the original fermion system. 
For example, the most popular condition ($s=N, N=\Omega, T(+)=T(-)=\hbar/4$) 
gives 
\beq\label{5-11}
M=(\hbar/2)\Omega \ . 
\end{equation}
As was mentioned in \S 2, the case (\ref{5-11}) shows that $\Omega$ 
fermions occupy fully in the lower level, if the interaction is 
switched off.

\section{Discussion}

Finally, in this discussion, we will sketch the results 
obtained in the frame of the $su(2)\otimes su(2)$-coherent 
state $\kket{c_0''}$, 
which is expressed as follows: 
\beq\label{6-1}
\kket{c_0''}=e^{-|c_+|^2}\exp\left(\sqrt{2c_+|c_+|}{\hat b}_+^*\right)
\cdot e^{-|c_-|^2}\exp\left(\sqrt{2c_-|c_-|}{\hat b}_-^*\right)\kket{0} \ .
\end{equation}
Here, $(c_\sigma, c_{\sigma}^*)$ plays the same role as that in the two cases 
we already discussed; 
the boson-type canonical variables. 
The state $\kket{c_0''}$ satisfies 
\beq\label{6-2}
{\hat b}_{\sigma}\kket{c_0''}=\sqrt{2c_{\sigma}|c_{\sigma}|}\kket{c_0''} \ . 
\qquad (\sigma=\pm)
\end{equation}
Differently from the previous two cases, we must pay a special attention 
on the present one. 
The expectation values of ${\wtilde T}_{\pm}(\sigma)$ are given by 
\beqn\label{6-3}
& &\bbra{c_0''}{\wtilde T}_+(\sigma)\kket{c_0''}
=\sqrt{\hbar}c_{\sigma}^*\sqrt{\hbar|c_{\sigma}|^2} \ , 
\nonumber\\
& &\bbra{c_0''}{\wtilde T}_-(\sigma)\kket{c_0''}
=\sqrt{\hbar|c_{\sigma}|^2}\cdot\sqrt{\hbar}c_{\sigma} \ . 
\eeqn
However, for $\bbra{c_0''}{\wtilde T}_0(\sigma)\kket{c_0''}$, formally, 
we have 
\beq\label{6-4}
\bbra{c_0''}{\wtilde T}_0(\sigma)\kket{c_0''}=\hbar/4+\hbar |c_{\sigma}|^2\ . 
\end{equation}
The problem concerning the expression is that we should include the term 
$\hbar/4$ into the expression (\ref{6-4}) or not. 
In Ref. \citen{14}, quantal fluctuation around the expectation value 
calculated by $\kket{c_0'}$ was discussed. 
If we follow the conclusion of Ref. \citen{14}, the term $\hbar/4$ 
should be included in the quantal fluctuations. 
Then, we should set up 
\beq\label{6-5}
\bbra{c_0''}{\wtilde T}_0(\sigma)\kket{c_0''}=\hbar |c_{\sigma}|^2\ . 
\end{equation}
However, for the calculation of the expectation value of the 
Hamiltonian ${\wtilde H}$, we can escape from this problem, because 
${\wtilde H}$ consists of the difference 
$(\hbar{\hat b}_+^*{\hat b}_+-\hbar{\hat b}_-^*{\hat b}_-)$ and the 
term $\hbar/4$ disappears. 

From the above argument, we have the Hamiltonian 
\beqn\label{6-6}
{H^0}''&=&\bbra{c_0''}{\wtilde H}\kket{c_0''} \nonumber\\
&=&\epsilon\cdot(\hbar c_+^*c_+-\hbar c_-^*c_-)\nonumber\\
& &-G\left(\sqrt{\hbar c_-^*c_-}\cdot \hbar c_+^*c_-\cdot 
\sqrt{\hbar c_+^*c_+} 
+\sqrt{\hbar c_+^*c_+}\cdot \hbar c_-^*c_+\cdot 
\sqrt{\hbar c_-^*c_-} \right) \ . 
\eeqn
If taking into account the relation $M=(\hbar/2)(c_+^*c_++c_-^*c_-)$, 
we have another form 
\beqn\label{6-7}
{H^0}''&=&-\epsilon M+2\epsilon\cdot \hbar c^*c \nonumber\\
& &-G\left(
\sqrt{\hbar}c^*\cdot\sqrt{\hbar c^*c}(2M-\hbar c^*c)
+(2M-\hbar c^*c)\sqrt{\hbar c^*c}\cdot\sqrt{\hbar}c\right) \ . 
\eeqn
As was done in Ref. \citen{12}, we introduce the canonical variable 
$(\psi, K)$ in the form 
\beq\label{6-8}
\sqrt{\hbar}c=\sqrt{K}e^{-i\psi} \ , \qquad 
\sqrt{\hbar}c^*=\sqrt{K}e^{i\psi} \ . 
\end{equation}
Then, ${H^0}''$ in the form (\ref{6-7}) can be rewritten as 
\beqn\label{6-9}
{H^0}''&=&-2M+2\epsilon K-2G\cdot K(2M-K)+4GK(2M-K)(\sin \psi/2)^2 \ .
\eeqn
On the other hand, ${H^0}'$ shown in the form (\ref{5-8}) is rewritten in 
the form 
\beqn\label{6-10}
{H^0}'&=&\epsilon(T(+)-T(-)-M)+2\epsilon K \nonumber\\
& &-2G\sqrt{K(2T(+)+K)}\cdot\sqrt{2M-K}\sqrt{2T(-)+2M-\hbar-K} \nonumber\\
& &-4G\sqrt{K(2T(+)+K)}\cdot\sqrt{2M-K}\sqrt{2T(-)+2M-\hbar-K}
(\sin \psi/2)^2 \ .\qquad
\eeqn
Both Hamiltonians are valid only in the region 
\beq\label{6-11}
0 \leq K \leq 2M-\hbar/2 \quad {\rm or}\quad 2M \ .
\end{equation}
The minimum point for both Hamiltonians (energies) appears at least at 
\beq\label{6-12}
\psi=0 \ .
\end{equation}
Therefore, it may be interesting to investigate at which points for 
$K$ the minimum points appear. 
For the Hamiltonian (the energy) (\ref{6-9}), the behavior near $K=0$ 
is determined by the factor $2(\epsilon-2GM)$. 
If $\epsilon > 2GM$, the energy increases in any region of $K$, 
and then, the minimum point appears at $K=0$. 
If $\epsilon < 2GM$, the energy decreases and afterward increases. 
Then, the minimum point appears at a certain value of $K$ ($>0$). 
This means that, at $\epsilon=2GM$, the phase change appears. 
In the case of the Hamiltonian (the energy) (\ref{6-10}), the behavior near 
$K=0$ is determined by the factor $2\epsilon K-2G\sqrt{K(2T(+)+K)}$. 
In this case, the energy decreases at any point of $K$ near $K=0$ and 
afterward increases. 
Therefore, the minimum point appears at a certain point ($>0$) except $K=0$ 
in the case $G=0$. 
From this argument, we cannot expect the sharp phase change. 
This is in the same situation as that in the case of the 
two-level pairing model.\cite{12} 
In the succeeding paper, we will repeat the above-mentioned point 
qualitatively including other various features.



\appendix

\section{Further interpretation of ${\hat X}_\nu^*$ and ${\hat Y}_s^*$ 
defined in the relation (\ref{2-8})}

We introduce the following operator: 
\beq\label{a1}
{\hat T}_l^*(n_1,\cdots ,n_{2l})={\hat c}_{n_1}^*(-)\cdots 
{\hat c}_{n_{2l}}^*(-)\ . \qquad (2l;{\hbox{\rm positive integer}})
\end{equation}
It may be clear from the relation (\ref{2-7}) that $({\hat c}_m^*(\pm))$ is 
spinor, and then, the operator (\ref{a1}) is the $(-l)$-th component 
of the tensor operator with rank $l$ for $({\hat {\cal S}}_{\pm,0})$ 
and we have 
\beq\label{a2}
[\ {\hat {\cal N}}\ , \ {\hat T}_{l}^*(n_1,\cdots ,n_{2l})\ ]
=2l{\hat T}_l^*(n_1,\cdots , n_{2l}) \ .
\end{equation}
The operator ${\hat Y}_s^*$ given in the relation (\ref{2-8b}) is 
nothing but 
\begin{equation}\label{a3}
{\hat Y}_s^*={\hat T}_s^*(m_1,\cdots ,m_{2s}) \ .
\end{equation}
It may be self-evident that ${\hat Y}_s^*$ satisfies the condition 
(\ref{2-6b}). 

The $m$-th component of the present tensor operator is expressed in the form 
\beq\label{a4}
{\hat T}_{lm}^*(n_1,\cdots , n_{2l})=
\sqrt{\frac{(l-m)!}{(2l)!(l+m)!}}\left({\vect {\cal S}}_+\right)^{l+m}
{\hat T}_l^*(n_1,\cdots ,n_{2l}) \ .
\end{equation}
Here, for any operators ${\hat A}$ and ${\hat B}$, $\vect{A}{\hat B}$ is 
defined as 
\beq\label{a5}
{\vect A}{\hat B}=[\ {\hat A}\ , \ {\hat B}\ ] \ . 
\end{equation}
With the use of the form (\ref{a4}), we can construct the scalar operator 
as follows: 
\beqn\label{a6}
& &
{\hat {\mib T}}_{00}^*(n_1,\cdots ,n_{2\lambda};n_1',\cdots , n_{2\lambda}')
=\sum_{\mu=-\lambda}^{\lambda}(-)^{\lambda-\mu}
{\hat T}_{\lambda\mu}^*(n_1,\cdots ,n_{2\lambda})
{\hat T}_{\lambda-\mu}^*(n_1',\cdots ,n_{2\lambda}') \ .\nonumber\\
& &
\eeqn
Clearly, 
${\hat {\mib T}}_{00}^*(n_1,\cdots ,n_{2\lambda};n_1',\cdots , n_{2\lambda}')$ 
commutes with ${\hat {\cal S}}_{\pm,0}$ and symbolically it is expressed in the form 
\beqn\label{a7}
& &
{\hat {\mib T}}_{00}^*(n_1,\cdots ,n_{2\lambda};n_1',\cdots , n_{2\lambda}')
\nonumber\\
&=&\sum_{(\rho,\rho')}D(n_1,\cdots ,n_{2\lambda};
n_1',\cdots ,n_{2\lambda}'|\rho_1,\cdots ,\rho_{2\lambda};\rho_1', \cdots ,
\rho_{2\lambda}') 
\nonumber\\
& &\qquad\times
{\hat c}_{\rho_1}^*(+)\cdots {\hat c}_{\rho_{2\lambda}}^*(+)
{\hat c}_{\rho_1'}^*(-)\cdots {\hat c}_{\rho_{2\lambda}'}^*(-) \ . 
\eeqn
Of course, we have 
\beq\label{a8}
{\hat {\mib T}}_{00}(n_1,\cdots ,n_{2\lambda};n_1',\cdots , n_{2\lambda}')
{\hat Y}_s^*\rket{0}=0 \ .
\end{equation}
From the above argument, ${\hat X}_\nu^*$ can be constructed in the 
form which satisfies 
$[{\hat{\cal N}},{\hat X}_\nu^*]$\break
$=2\nu{{\hat X}_\nu}^*$.

\section{The Holstein-Primakoff boson representation of the $su(1,1)$-algebra 
for boson pair and its classical counterpart}

In this Appendix, we summarize the $su(1,1)$-algebra for boson pair 
which is a base of the $su(1,1)\otimes su(1,1)$-algebra for boson pairs 
appearing in the text of this paper. 
In the present case, the generators are given in the form 
\beq\label{b1}
{\hat \tau}_+=({\hat b}^*)^2/2\ , \qquad
{\hat \tau}_-=({\hat b})^2/2\ , \qquad
{\hat \tau}_0=1/4+({\hat b}^*{\hat b})/2\ .
\end{equation}
Here, $({\hat b},{\hat b}^*)$ denotes boson operator, and for simplicity, 
$\hbar=1$ is taken. 
The state with the minimum weight $\kket{t}$ is obtained under the condition 
\beq\label{b2}
{\hat \tau}_-\kket{t}=0\ , \qquad 
{\hat \tau}_0\kket{t}=t\kket{t} \ . \quad (t=1/4, 3/4)
\end{equation}
For $t$, we have the form 
\beq\label{b3}
\kket{t}=\kket{0}\quad{\rm for}\quad t=1/4\ , \qquad
\kket{t}={\hat b}^*\kket{0}\quad{\rm for}\quad t=3/4\ .
\end{equation}
Here, $\kket{0}$ denotes the vacuum for $({\hat b},{\hat b}^*)$. 
The state $\kket{t}$ is expressed as 
\beq\label{b4}
\kket{t}=({\hat b}^*)^{2t-1/2}\kket{0} \ . 
\end{equation}
Then, by denoting the eigenvalue of ${\hat \tau}_0$ as 
$k+t$ $(k=0, 1,2,\cdots)$, the orthogonal set $\{\kket{k;t}\}$ is expressed 
as follows: 
\beqn\label{b5}
\kket{k;t}&=&2^k\left(\sqrt{(2k+2t-1/2)!}\right)^{-1}
({\hat \tau}_+)^k\kket{t} \nonumber\\
&=&\left(\sqrt{(2k+2t-1/2)!}\right)^{-1}
({\hat b}^*)^{2k+2t-1/2}\kket{0} \ . 
\eeqn

In order to obtain the Holstein-Primakoff representation, 
we adopt the MYT mapping method.\cite{11} 
We prepare a new boson space spanned by boson $({\hat c},{\hat c}^*)$: 
\beq\label{b6}
\ket{k}=\left(\sqrt{k!}\right)^{-1}({\hat c}^*)^k\ket{0}\ . 
\quad ({\hat c}\ket{0}=0)
\end{equation}
Then, the mapping operator ${\maru U}$ is defined as 
\beq\label{b7}
{\maru U}=\sum_{k=0}^{\infty}\ket{k}\bbra{k;t}\ . 
\end{equation}
With the use of ${\maru U}$, we get the Holstein-Primakoff boson 
representation in the form 
\beqn\label{b8}
& &{\maru \tau}_+={\maru U}{\hat \tau}_+{\maru U}^{\dagger}
={\hat c}^*\cdot\sqrt{2t+{\hat c}^*{\hat c}} \ , \nonumber\\
& &{\maru \tau}_-={\maru U}{\hat \tau}_-{\maru U}^{\dagger}
=\sqrt{2t+{\hat c}^*{\hat c}}\cdot{\hat c} \ , \nonumber\\
& &{\maru \tau}_0={\maru U}{\hat \tau}_0{\maru U}^{\dagger}
=t+{\hat c}^*{\hat c} \ .
\eeqn
Here, $t$ denotes 
\beq\label{b9}
t=1/4\ , \quad 3/4\ .
\end{equation}
The form (\ref{b8}) is also valid for the case $t=1/2,1,3/2, \cdots$, 
which is treated in (I). 
The above is our first interest. 

As a second interest, we investigate the classical counterpart of the 
form (\ref{b1}) or (\ref{b8}). 
For this purpose, we, first, introduce the following state: 
\beqn\label{b10}
\kket{c_0}&=&\left(\sqrt{U}\right)^{-1}
\exp\left(\frac{V}{U}{\hat \tau}_+\right)\kket{0} \nonumber\\
&=&\left(\sqrt{U}\right)^{-1}
\exp\left(\frac{V}{2U}({\hat b}^*)^2\right)\kket{0} \ .
\eeqn
Here, $V$ denotes a complex parameter and $U$ is given as 
\beq\label{b11}
U=\sqrt{1+|V|^2} \ .
\end{equation}
The state $\kket{c_0}$ is a vacuum of the boson operator 
$({\hat \beta},{\hat \beta}^*)$ defined as 
\bsub\label{b12}
\beqn
& &{\hat \beta}=U{\hat b}-V{\hat b}^* \ , \nonumber\\
& &{\hat \beta}^*=-V^*{\hat b}+U{\hat b}^* \ , 
\label{b12a}\\
{\rm i.e.,}\qquad
& &{\hat b}=U{\hat \beta}+V{\hat \beta}^*\ , \nonumber\\
& &{\hat b}^*=V^*{\hat \beta}+U{\hat \beta}^*\ .
\label{b12b}
\eeqn
\esub
In relation to $\kket{c_0}$, we introduce the state $\kket{c^0}$ in the form 
\beq\label{b13}
\kket{c^0}=\left(\sqrt{3/2-2t}+\sqrt{2t-1/2}{\hat \beta}^*\right)
\kket{c_0} \ . \quad (t=1/4, 3/4)
\end{equation}
Of course, we have $(\sqrt{3/2-2t})^2+(\sqrt{2t-1/2})^2=1$ and 
$\sqrt{3/2-2t}\cdot\sqrt{2t-1/2}=0$. 
With the use of the state $\kket{c^0}$, we have the relations 
\beqn
& &\tau_+^0=\bbra{c^0}{\hat \tau}_+\kket{c^0}=2tUV^* \ , \nonumber\\
& &\tau_-^0=\bbra{c^0}{\hat \tau}_-\kket{c^0}=2tUV \ , \nonumber\\
& &\tau_0^0=\bbra{c^0}{\hat \tau}_0\kket{c^0}=t+2t|V|^2 \ , 
\label{b14}\\
& &\bbra{c}\frac{\partial}{\partial z}\kket{c}=\frac{1}{2}\cdot 
2t\left(V^*\frac{\partial V}{\partial z}-V\frac{\partial V^*}{\partial z}
\right)\ . 
\label{b15}
\eeqn
Instead of $(V,V^*)$, we introduce new parameter $(c,c^*)$ obeying 
\beq\label{b16}
\bbra{c^0}\frac{\partial}{\partial c}\kket{c^0}=c^*/2\ , \qquad
\bbra{c^0}\frac{\partial}{\partial c^*}\kket{c^0}=-c/2\ .
\end{equation}
With the use of the relation (\ref{b15}), $(V,V^*)$ can be 
expressed as 
\beq\label{b17}
V=c/\sqrt{2t} \ , \qquad V^*=c^*/\sqrt{2t} \ . 
\end{equation}
The relation (\ref{b16}) tells that $(c,c^*)$ plays a role of 
canonical variable in boson type in classical mechanics. 
Then, we have the following form for $(\tau_{\pm,0}^0)$: 
\beq\label{b18}
\tau_+^0=c^*\sqrt{2t+c^*c}\ , \qquad 
\tau_-^0=\sqrt{2t+c^*c}\ c \ , \qquad
\tau_0^0=t+c^*c\ . 
\end{equation}
We can see that the form (\ref{b18}) is a classical counterpart of the form 
(\ref{b8}), i.e., (\ref{b1}) under the replacement
\beq\label{b19}
{\hat c}\longrightarrow c\ , \qquad {\hat c}^*\longrightarrow c^*\ . 
\end{equation}
Further, we can show that the Poisson bracket for the expression 
(\ref{b18}) gives us the same form as the commutation relation in 
Dirac's sense.

Our third interest is related to the state 
\beq\label{b20}
\kket{c_0'}=\left(\sqrt{U}\right)^{-1}
\exp\left(\frac{\gamma}{|\gamma|}\frac{V}{2U}({\hat b}^*)^2
+\frac{\sqrt{\gamma}}{U}{\hat b}^*\right)\kket{0} \ .
\end{equation}
Here, $\gamma$ is a complex parameter and $V$ is real and $U$ is given by 
\beq\label{b21}
U=\sqrt{1+V^2} \ .
\end{equation}
The state $\kket{c_0'}$ is a vacuum for the boson operator 
$({\hat \beta},{\hat \beta}^*)$ defined as 
\bsub\label{b22}
\beqn
& &{\hat \beta}=U{\hat b}-\frac{\gamma}{|\gamma|}V{\hat b}^*-\sqrt{\gamma} 
\ , \nonumber\\
& &{\hat \beta}^*=-\frac{\gamma^*}{|\gamma|}V{\hat b}+U{\hat b}^*
-\sqrt{\gamma^*} \ , 
\label{b22a}\\
{\rm i.e.,}\qquad
& &{\hat b}=U{\hat \beta}+\frac{\gamma}{|\gamma|}V{\hat \beta}^*
+\sqrt{\gamma}(U+V)\ , \nonumber\\
& &{\hat b}^*=\frac{\gamma^*}{|\gamma|}V{\hat \beta}+U{\hat \beta}^*
+\sqrt{\gamma^*}(U+V)\ .
\label{b22b}
\eeqn
\esub
In relation to the state $\kket{{c_0}'}$, we introduce the state 
$\kket{{c^0}'}$ 
in the form 
\beq\label{b23}
\kket{{c^0}'}=\left(\sqrt{3/2-2t}+\sqrt{2t-1/2}{\hat \beta}^*\right)
\kket{{c_0}'} \ . \qquad
(t=1/4,3/4)
\end{equation}
Of course, we have $(\sqrt{3/2-2t})^2+(\sqrt{2t-1/2})^2=1$ 
and $\sqrt{3/2-2t}\cdot\sqrt{2t-1/2}=0$. 
The state $\kket{{c^0}'}$ gives the following relations: 
\beqn
& &{\tau_+^0}'=\bbra{{c^0}'}{\hat \tau}_+\kket{{c^0}'}=
2t\frac{\gamma^*}{|\gamma|}UV+\gamma^*(U+V)^2 \ , 
\nonumber\\
& &{\tau_-^0}'=\bbra{{c^0}'}{\hat \tau}_-\kket{{c^0}'}=
2t\frac{\gamma}{|\gamma|}UV+\gamma(U+V)^2 \ , 
\nonumber\\
& &{\tau_0^0}'=\bbra{{c^0}'}{\hat \tau}_0\kket{{c^0}'}=
t+2tV^2+\frac{1}{2}|\gamma|(U+V)^2 \ , 
\label{b24}\\
& &\bbra{{c^0}'}\frac{\partial}{\partial z}\kket{{c^0}'}=
\frac{1}{2}\left(2t\frac{V^2}{|\gamma|^2}+\frac{1}{2|\gamma|}(U+V)^2
\right)\left(\gamma^*\frac{\partial \gamma}{\partial z}-
\gamma\frac{\partial \gamma^*}{\partial z}\right) \ . 
\label{b25}
\eeqn
Instead of $(\gamma,\gamma^*)$, we introduce new parameter $(c,c^*)$ obeying 
\beq\label{b26}
\bbra{{c^0}'}\frac{\partial}{\partial c}\kket{{c^0}'}=c^*/2\ , \qquad
\bbra{{c^0}'}\frac{\partial}{\partial c^*}\kket{{c^0}'}=-c/2\ . 
\end{equation}
The relation (\ref{b26}) with (\ref{b25}) gives us 
\beq\label{b27}
\gamma=\frac{c}{|c|}\cdot\frac{2(|c|^2-2tV^2)}{(U+V)^2} \ , \qquad
\gamma^*=\frac{c^*}{|c|}\cdot\frac{2(|c|^2-2tV^2)}{(U+V)^2} \ . 
\end{equation}
In the same meaning as that in the previous case, $(c,c^*)$ plays a role of 
the canonical variable in classical mechanics. 
Then, the form (\ref{b24}) is expressed in the following form: 
\beqn\label{b28}
& &{\tau_+^0}'=\frac{c^*}{|c|}(|c|^2+2tV(U-V)) \ , \nonumber\\
& &{\tau_-^0}'=\frac{c}{|c|}(|c|^2+2tV(U-V)) \ , \nonumber\\
& &{\tau_0^0}'=t+|c|^2 \ . 
\eeqn
By changing $V$ in various forms as a function of $|c|^2$, we get various 
forms for $({\tau_{\pm,0}^0}')$. 
For instance, under the form (\ref{b28}), we have 
\beq\label{b29}
{\tau_-^0}'=\frac{c}{|c|}\sqrt{|c|^2(2t-1+|c|^2)}
=\sqrt{2t-1+|c|^2}\cdot c \ .
\end{equation}
The condition (\ref{b26}) gives us 
\beq\label{b30}
V=\sqrt{\frac{|c|^2}{2t}}\frac{|2t-1|}
{\sqrt{(2t-1)^2+(\sqrt{2t-1+|c|^2}+|c|^2)^2}} \ . 
\end{equation}
This form is used in \S 5. 
Generally, we have the following form: 
\beqn
& &\tau_-=\sqrt{2\tau+|c|^2}\cdot c \ , 
\label{b31}\\
& &V=\sqrt{\frac{|c|^2}{2t}}\frac{|2\tau|}
{\sqrt{(2\tau)^2+2(t-\tau)(\sqrt{2\tau+|c|^2}+|c|^2)^2}} \ . 
\label{b32}
\eeqn
Various boson-pair coherent states were investigated in Ref. \citen{15} 
under the viewpoint of the deformed boson scheme.

\end{document}